\documentstyle[12pt,epsfig,subfigure,amsfonts]{article}

\newcommand{\be}{\begin{equation}}
\newcommand{\ee}{\end{equation}}
\newcommand{\beq}{\begin{equation}}
\newcommand{\eeq}{\end{equation}}
\newcommand{\ba}{\begin{eqnarray}}
\newcommand{\ea}{\end{eqnarray}}
\newcommand{\bea}{\begin{eqnarray}}
\newcommand{\eea}{\end{eqnarray}}

\begin{document}
\baselineskip=15.5pt \pagestyle{plain} \setcounter{page}{1}
\begin{titlepage}

\vskip 0.8cm

\begin{center}

{\LARGE Dynamics of holographic thermalization} \vskip .3cm

\vskip 1.cm

{\large {Walter Baron$^{a,}${\footnote{\tt
wbaron@fisica.unlp.edu.ar}}, Dami\'an Galante$^{b,
c,}${\footnote{\tt dgalante@perimeterinstitute.ca}} and Martin
Schvellinger$^{a,}${\footnote{\tt martin@fisica.unlp.edu.ar}}}}

\vskip 1.cm

$^{a}${\it IFLP-CCT-La Plata, CONICET and Departamento  de
F\'{\i}sica,  Universidad Nacional de La Plata.  Calle 49 y 115,
C.C. 67, (1900) La Plata,  Buenos Aires,
Argentina.} \\

\vskip 0.5cm

$^{b}${\it Perimeter Institute for Theoretical Physics, Waterloo,
Ontario N2J 2W9, Canada.} \\

\vskip 0.5cm

$^{c}${\it Department of Applied Mathematics, University of Western
Ontario, \\  London, Ontario N6A 5B7, Canada.}

\vspace{1.cm}

{\bf Abstract}

\end{center}

Dynamical evolution of thin shells composed by different kinds of
degrees of freedom collapsing within asymptotically AdS spaces is
explored with the aim of investigating models of holographic
thermalization of strongly coupled systems. From the quantum field
theory point of view this corresponds to considering different
thermal quenches. We carry out a general study of the thermalization
time scale using different parameters and space-time dimensions, by
calculating renormalized space-like geodesic lengths and rectangular
minimal area surfaces as extended probes of thermalization, which
are dual to two-point functions and rectangular Wilson loops.
Different kinds of degrees of freedom in the shell are described by
their corresponding equations of state. We consider a scalar field,
as well as relativistic matter, a pressureless massive fluid and
conformal matter, which can be compared with the collapse of an
AdS-Vaidya thin shell. Remarkably, in the case of AdS$_5$, for
conformal matter, the thermalization time scale becomes much larger
than the others. Furthermore, in each case we also investigate
models where the cosmological constants of the inner and outer
regions separated by the shell are different. We found that in this
case only a scalar field shell collapses, and that the
thermalization time scale is also much larger than the AdS-Vaidya
case.

\noindent

\end{titlepage}

\newpage

\tableofcontents

\vfill

\newpage

\section{Introduction and motivation}

The idea of the present work is to investigate different kinds of
consistent holographic thermal quenches modeling thermalization
processes in strongly coupled systems. As we shall explain in
detail, by construction, they satisfy the general relativity
equations of motion and the positive energy conditions. Our
particular interest is focused on the strongly coupled quark-gluon
plasma (QGP) produced by the collision of heavy ions at the
Relativistic Heavy Ion Collider (RHIC) and the Large Hadron Collider
(LHC). As it is well-known the formation and evolution of a
quark-gluon plasma can be viewed as a sequence four distinct steps.
First, two heavy ions, typically gold nuclei, move towards each
other at relativistic velocities, having kinetic energies of order
100 GeV/nucleon. Next, an almond-shape region where the two nuclei
collide is developed, and a part of their kinetic energy transforms
into intense heat, leading to the beginning of formation of the
plasma of quarks and gluons. This is what has been called
thermalization of the plasma. When the thermalization is completed
the resulting system is a strongly coupled QGP. After a very short
while the system expands, cools down and, finally, a multitude of
hadrons emerges from the plasma. Given the fact that the QGP in
thermal equilibrium behaves as an strongly coupled system, a
reasonable working hypothesis is that the thermalization described
above may also occur within a strongly coupled regime of QCD. In a
number of interesting articles thermalization has been addressed
using the gauge/gravity duality
\cite{Maldacena:1997re,Gubser:1998bc,Witten:1998qj}. In these papers
the dual process is modeled as the collapse of a thin shell moving
at the speed of light, using an AdS-Vaidya type metric, which
represents a thermal quench
\cite{Das:2010yw,AbajoArrastia:2010yt,Albash:2010mv,Ebrahim:2010ra,
Balasubramanian:2010ce,Balasubramanian:2011ur,Garfinkle:2011hm,
Aparicio:2011zy,Allais:2011ys,Keranen:2011xs,
Garfinkle:2011tc,Das:2011nk,Hubeny:2012ry,Galante:2012pv,Caceres:2012em,
Wu:2012rib}. Interestingly, there have also been studies on
holographic thermalization described as a dual process of black hole
formation
\cite{Danielsson:1999fa,Giddings:1999zu,Janik:2006gp,Janik:2006ft,
Chesler:2008hg,Chesler:2009cy,Bhattacharyya:2009uu,Lin:2008rw}. In
addition, to our knowledge, reference \cite{Shuryak:2005ia} has been
the first one to consider a gravity dual picture of the entire
process of strongly coupled supersymmetric Yang-Mills (SYM) plasma
formation and cooling using a model where the scattering process
initially creates a holographic shower in the AdS bulk. It has been
argued \cite{Shuryak:2005ia} that the subsequent gravitational fall
leads to a moving black hole, which is the gravity dual model
corresponding to an expanding and cooling heavy-ion fireball.
Moreover, very recently, it has been investigated the high and low
temperature behavior of non-local observables in strongly coupled
gauge theories that are dual to AdS space-time
\cite{Fischler:2012ca}.

In the present case we shall consider a holographic dual description
of the thermalization process. From the point of view of the
boundary quantum field theory (QFT), the initial state that one
considers is a system at zero temperature. Then, there is a sudden
injection of energy which induces an abrupt change in the state of
the system. The system evolves leading to a final thermal state
which will be a strongly coupled SYM plasma. A very important
question is how to model the thermal evolution of the system from
the zero temperature state towards the thermally equilibrated SYM
plasma, keeping in mind that the initial condition is a thermal
quench instead of an adiabatic change. Indeed, this is a very hard
problem if one tries to study its dynamics in terms of QFT methods.
On the other hand, the holographic evolution of a thermal quench can
be easily followed by numerical calculations in its gravity dual
model. So far, the studies \cite{Das:2010yw}-\cite{Wu:2012rib} have
considered the evolution of an AdS-Vaidya thin shell, even though at
the moment it is not known how to get the initial Vaidya shell
condition from a QFT evolution. On the gravity side, we know that it
is also possible to solve the equations of motion (EOM) of shells
composed by different degrees of freedom, {\it i.e.} whose dynamics
is described by different equations of state (EOS). Depending on the
particular EOS the shells will move and collapse at different
velocities in the bulk. This is very interesting since it allows us
to investigate the thermalization time scale of different kind of
shells. On the boundary theory side, after the collision occurs, the
system evolves in a certain way until it reaches thermal
equilibrium. On the other hand, on the holographic gravitational
dual model, this should be reflected on the evolution of a
collapsing shell, which depends on the EOS governing the degrees of
freedom which compose it. Thus, we shall be focused at investigating
the variation of the thermalization time scale of two-point
functions of gauge invariant local operators and rectangular Wilson
loops, by calculating their dual renormalized space-like geodesic
lengths and rectangular minimal area surfaces. These are extended
probes within the dual geometry for thermal equilibrium in the dual
QFT. The interesting new feature of this work is that we change the
nature of the shell composition. We study different kinds of degrees
of freedom in the shell which are described by distinct EOS. These
include a scalar field, conformal matter, relativistic matter and a
pressureless massive fluid, which can be compared with the collapse
of an AdS-Vaidya thin shell. Furthermore, we also investigate models
where the cosmological constants of the inner and outer regions are
different. On the field theory side this corresponds to changes in
the coupling of a SYM theory at zero temperature compared with the
SYM plasma coupling at thermal equilibrium.

It is worth noting that for a SYM plasma in thermal equilibrium the
system is probed at momentum scales below the equilibrium
temperature $T$. This is the so-called hydrodynamical regime, at
which the gauge/string duality has been proved to be particularly
useful. An important number of investigations have been done in this
framework. A very important work in this context is the one of
reference \cite{Policastro:2001yc}, where it has been calculated the
shear viscosity of the finite-temperature ${\cal {N}} = 4$ $SU(N)$
SYM theory plasma, in the large $N$ limit, at the strong-coupling
regime. The first leading order string theory corrections to the
shear viscosity over entropy density ratio of strongly coupled SYM
plasmas has been obtained in \cite{Buchel:2004di}. Besides,
electrical charge transport coefficients of strongly coupled SYM
plasmas have also been investigated within the gauge/string duality.
These include the electrical conductivity, which in the large
coupling limit was firstly calculated in \cite{Policastro:2002se},
while finite 't Hooft coupling corrections were obtained from type
IIB string theory corrections at order $\alpha'^3$ in
\cite{Hassanain:2010fv,Hassanain:2011fn}. Additionally, the
photoemission rates of this plasma have been computed in
\cite{CaronHuot:2006te} using the gauge/string duality, while the
corresponding leading order string theory corrections have been
reported more recently in \cite{Hassanain:2011ce,Hassanain:2012uj}.
Besides, holographic photon and dilepton production in a
thermalizing plasma have been investigated within the quasi-static
approximation \cite{Baier:2012tc,Baier:2012ax,Steineder:2012si}.

We would like to emphasize some interesting conclusions which follow
from our numerical results. We observe that from the curves of
thermalization discussed in this paper, by fixing to one both the
inner and outer radii, the shells composed by a pressureless massive
fluid and by a scalar field are very close to each other, and almost
overlap completely the curve corresponding to the AdS-Vaidya shell
moving at the speed of light. On the other hand, relativistic matter
thermalizes later, depending on its EOS the difference becomes more
important, and finally the shell composed by conformal matter
thermalizes much later than the time the AdS-Vaidya shell takes to
collapse. We would like to emphasize that this large thermalization
time delay is a very remarkable effect, since we think that it opens
the possibility of developing new type of models showing slower
thermalization in comparison with the AdS-Vaidya models. This is for
both space-like renormalized geodesic lengths for space-time
dimensions $d=2$, 3 and 4, and for renormalized rectangular minimal
area surfaces for $d=3$ and 4. We also have numerically investigated
what happens when both radii are equal to each other but we change
both simultaneously.  Then, we study the effect on the
thermalization curves when the inner and outer radii are different.
We have obtained an interesting analytical result, namely: the
positive energy condition implies that the inner radius must be
equal or smaller than the outer one, {\it i.e.} the absolute value
of the vacuum energy density of the inner region must be equal or
larger than the one of the outer region. In addition, only in the
case of a shell composed by a scalar field the positive energy
condition allows for the collapse of the shell separating regions
with different inner and outer vacuum energy density to be produced.

This paper is organized as follows. In section 2 we introduce the
formalism, including the description of the thermal quenches
corresponding to different kinds of degrees of freedom living in the
collapsing shells. We derive the expressions for the velocity of the
shell and its mass function. Then, we describe the strongly coupled
SYM plasma thermalization process in terms of the evolution of a
massive shell. We introduce the extensive gravitational probes we
use to measure the thermalization time scale, which includes
renormalized space-like geodesic lengths and rectangular minimal
area surfaces. The latter correspond to Wilson loops in a
4-dimensional QFT on the boundary, and are proportional to
entanglement entropy for a 3-dimensional boundary QFT. These are
introduced in section 3. In section 4 we present our results on
holographic thermalization for different dimensions of space-time
and by exploring different sets of parameters. In the last section
we discuss the results.

\section{Thermal quenches and equations of state}

The dynamics of a massive thin shell is determined by the Israel
junction conditions \cite{Israel:1966rt}. The shell separates two
different geometries, each one being a solution of the Einstein
equations, and the Israel's conditions tell us how to match them.

Since the shell is massive, the inner solution will typically be a
vacuum one, while the outer geometry will be described by an
AdS-Schwarzschild-type solution. The shell can be made of ordinary
particles, like null dust as described by the AdS-Vaidya solution,
which gives for instance the geometry generated by a spherically
symmetric beam of photons in the Eikonal approximation
\cite{vaidya1,vaidya2}; by conformal matter as described in
\cite{Erdmenger:2012xu}; moreover, it can also be interpreted as the
domain wall of a solitonic solution connecting the inner and outer
geometries with different cosmological constants associated with the
vacuum expectation value of certain scalar field.

In the next section we will obtain the expression for the velocity
of the shell collapsing in a AdS$_{d+1}$ space-time. The inner
geometry will be a pure AdS space, while the outer space will be an
asymptotically AdS-Schwarzschild black hole. Notice though that we
allow for the radii of both anti-de Sitter spaces to be in principle
arbitrary. Moreover, in general terms its evolution can be followed
for any EOS governing the degrees of freedom of the shell. Thus, by
setting a particular EOS one can determine the velocity of the
shell.

\subsection{Shell velocity}

We find useful to describe the AdS spaces by using
Eddington-Finkelstein-like coordinates\footnote{The time coordinate
is defined as usual through $dt=dv+ R_f f^{-1}_{out}(z) dz$.}. In terms
of these the metrics inside and outside the shell are given
respectively by
\bea
ds^2_{in} &=& g^{(in)}_{MN} \, dX_0^M dX_0^N =
\frac{1}{{z}^2}\left(- dv^2-2 R_0 \, dv dz + d{\vec{x}}^2\right)
\, ,\cr ds^2_{out} &=& g^{(out)}_{MN} \, dX_f^M dX_f^N =
\frac{1}{{z}^2}\left(-f_{out}(z) \, dv^2-2 R_f \, dv dz +
d{\vec{x}}^2\right) \, ,
\eea
where indices $M$ and $N=1, \cdots, d+1$, while ${\vec {x}}=\{ x^i
\}$ with $i=1, \cdots, d-1$ and $R_0$ and $R_f$ are the AdS radii
corresponding to the inner and outer regions, respectively, and
\bea
f_{out}(z)=1-2M (R_f z)^d \, .
\eea
Given the above metrics inside and outside the shell, it is natural
to use the following shell embedding metric
\bea
ds^2_{shell} &=& h^{(shell)}_{\mu\nu} \, dx^\mu dx^\nu =
\frac{1}{z^2}\left(- d\tau^2+ d{\vec{x}}^2\right)  \, ,
\label{gshell}
\eea
where we have defined $x^0 \equiv \tau$, so that $h_{\mu\nu}$ is
conformally flat. The proper area of the shell allows us to identify
the $z$ variable inside, over and outside the shell.

The energy-momentum tensor necessarily has the form,
\bea
T_{MN}=\delta(\eta) \, S_{MN} - \rho_0 \, g^{(in)}_{MN} \,
\Theta(-\eta) - \rho_f \, g^{(out)}_{MN} \, \Theta(\eta) \, ,
\eea
where $\eta$ is the coordinate orthogonal to the shell in the
Gaussian normal coordinate system. $S_{MN}$ represents the
energy-momentum tensor of the shell. $\rho_0$  and $\rho_f$ denote
the vacuum energy density of the anti-de Sitter spaces,
$\rho_{0,f}=\frac{-d(d-1)}{2\kappa~ R_{0,f}^2}$, with $\kappa=8\pi
G,$ $G$ being the $(d+1)$-dimensional Newton constant.

By computing the divergence of the energy-momentum tensor,
$T^{MN}{}_{;N}$, and demanding the coefficients of $\delta(\eta)$
and $\delta'(\eta)$ to vanish separately, it can be shown that the
surface energy-momentum tensor must vanish in the normal directions,
$S^{M\eta}=0$, and the non-trivial components are conserved in the
lower-dimensional sense, {\it i.e.}
\bea
S^{\mu\nu}{}_{|\nu}=0 \, , \label{ECShell}
\eea
where $``\,|\,"$ denotes the covariant derivative constructed with
$h^{(shell)}_{\mu\nu}$. Another consequence of $T^{MN}{}_{;N}=0$ is
the junction condition
\bea
\{K_{\mu\nu}\} S^{\mu\nu}=\rho_0-\rho_f \, ,
\label{KS}\label{juncond2}
\eea
where $\{K_{\mu\nu}\} = \frac12\left[K_{\mu\nu}(in) +
K_{\mu\nu}(out)\right]$, while $K_{\mu\nu}=n_{\mu;\nu}$ denotes the
extrinsic curvature, $n$ being the normal vector to the shell.

We will consider a shell composed by a perfect fluid, so that
\bea
S^{\mu\nu}=z(\tau)^2(\epsilon+p)~u^\mu u^\nu + p~ h^{\mu\nu} \, ,
\eea
where the velocities $u^\mu$ are defined as $\frac{dx^\mu}{d\tau}$,
with $\tau$ being the conformal time, not to be confused with the
proper time. Then, equation (\ref{ECShell}) implies
\bea
\dot{\epsilon}=(d-1)\frac{\dot z}{z}(\epsilon + p)~,\label{ep.eq}
\eea
where dot stands for derivative with respect to $\tau$. In the above
expression $\epsilon$ is the energy density and $p$ is the pressure
within the shell.

Einstein equations of the $(d+1)-$dimensional space-time lead to the
so-called Israel junction conditions, namely:
\bea
[K_{\mu\nu}-h^{(shell)}_{\mu\nu} \, tr K]=\kappa \, S_{\mu\nu} \, ,
\label{IJC}
\eea
the square bracket $[\,\cdot\,]$ denotes the difference of the
quantity inside and outside the shell.

The velocity of the fluid is set to be in the radial direction, so
that $u^\mu \rightarrow (\dot v,\dot z,\vec0)$, and the normal
vector $n^\mu$ defined as the unit vector orthogonal to $u^\mu$ is
easily found and leads to the following extrinsic curvature
\bea
K_{x^i x^j}(out) &=& \frac{\sqrt{f_{out}+R_f^2~\dot{z}^2}}{R_f~
z^2}\,\delta_{ij}\, ,~~~i,j=1,2,\cdots,d-1~,\cr\cr
K_{\tau\tau}(out) &=& \frac{d}{dz}\frac{\sqrt{f_{out} +
R_f^2~\dot{z}^2}}{R_f~ z} \, ,
\eea
in the outer region. In the inner region there are similar
expressions by just replacing $f_{out} \Rightarrow 1$ and $R_f
\Rightarrow R_0$. Then, the Israel junction conditions become
\bea
\sqrt{R_0^{-2}+\dot z^2}-\sqrt{f_{out}~R_f^{-2}+\dot
z^2}=\frac{\kappa}{d-1}~\epsilon \, . \label{diff.eq}
\eea
After some algebra equation (\ref{diff.eq}) leads to
\bea
\sqrt{R_0^{-2} + {\dot z}^2} + \sqrt{f_{out}\; R_f^{-2} + {\dot
z}^2}=\frac{d-1}{\kappa \,
\epsilon}\left(R_0^{-2}-f_{out}~R_f^{-2}\right) \, , \label{sum.eq2}
\eea
which represents the junction condition (\ref{juncond2}). In fact,
it implies
\bea
\epsilon ~\frac{d\left( z \{K_{x^1 x^1}\}\right)}{dz } + (d-1)\, p
~\{K_{x^1x^1}\}=-\frac{d(d-1)}{2\kappa}\left(\frac1{R_0^2}-\frac1{R_f^2}\right)
\, . \label{KS.eq}
\eea
Then, using equation (\ref{ep.eq}) this differential equation can be
integrated to obtain equation (\ref{sum.eq2}).

Equations (\ref{diff.eq}) and (\ref{sum.eq2}) can be used to derive
the following expression
\bea
\dot z^2=
\frac{h^2-2\left(R_0^{-2}+f_{out}~R_f^{-2}\right)h+\left(R_0^{-2}-
f_{out}~R_f^{-2}\right)^2}{4h} \, ,\label{zdot2}
\eea
where in order to make the notation simpler we have introduced
$h=(\frac{\kappa~ \epsilon}{d-1})^2$.

Since we are assuming that the shell is composed by a perfect fluid
the entropy must be a constant, which can be negligibly small such
that its EOS can be reduced to $p=p(\epsilon)$. In many physical
situations the EOS can exactly or at least approximately be recast
in the form $p=a \, \epsilon$, with $a$ being a constant. For
instance, when $a=\frac1{d-1}$ it represents a fluid composed by
conformal matter ({\it i.e.} its degrees of freedom have a traceless
energy-momentum tensor). On the other hand, the case with $a=0$
corresponds to dust, while $a=-1$ (see \cite{Guth:1987}) can be
modeled by a scalar field \footnote{To see this, notice that 
the energy-momentum tensor for a scalar field is
\bea
T_{\mu\nu}=\partial_\mu \phi\partial_\nu
\phi-g_{\mu\nu}\left[\frac12
\partial^\alpha\phi\partial_\alpha\phi+V(\phi)\right] \,, \nonumber
\eea
so, using that a the embedding metric of a hypersurface with normal vector $n_\mu$ is given by $h^{shell}_{\mu\nu}=g_{\mu\nu}-n_\mu n_\nu$ and the fact that in the thin-shell approximation $\partial_\mu \phi\propto n_\mu$ at 
leading order, $T_{\mu\nu}$ decompose in the sum of two terms, one proportional to $h_{\mu\nu}$ and the other one to $n_\mu n_{\nu}$. 
On the other hand if we require the shell to be composed by a perfect fluid, the energy momentum tensor must decomposes in a factor proportional to $h_{\mu\nu}$ and a factor proportional to $U_\mu U_\nu$, with $U_\mu\bot n_\mu$ being the four-velocity of the fluid. Then the energy momentum tensor is forced to be proportional to the metric, which means $p=-\epsilon$.}. In cosmological applications these are commonly
employed in order to describe the radiation, matter and dark energy
dominated eras. We will consider all these situations along this
paper and, in addition, we will include the case of relativistic
matter. With the purpose of illustrating this situation we will take
a particular example where $a=\frac9{10}\frac1{d-1}$ for
relativistic matter.

Using this equation of state, equation (\ref{ep.eq}) leads to the
energy density
\bea
\epsilon=\epsilon_0 \, z(\tau)^{A} \, ,
\eea
where $A=(d-1)(a+1)$ and $\epsilon_0$ is set by the initial
conditions. For instance, we will fix it by demanding that the shell
is at rest at a given position $z=z_0$.\footnote{Of course, the
situation is different for a massless dust fluid. In this case the
shell moves at the speed of light, and obviously it cannot be set at
rest at any position. Nevertheless, it can be considered as a
limiting case with $\dot z\rightarrow\infty$.} Therefore, one may
write
\bea
h(z)&=&\left(R_0^{-1}- \sqrt{f_{out}(z_0)}~R_f^{-1}\right)^2 \,
\left(\frac{z}{z_0}\right)^{2A} \, .
\eea
Notice that since the cut-off $z_0$ can be arbitrarily small, the
weak energy condition requires
\bea
R_f\geq R_0 \, .
\eea
Next, we derive the mass function of shells.

\subsection{Mass function}

Strictly speaking the analysis above corresponds to a shell of zero
thickness. Nevertheless, for computational purposes we will consider
the limiting case of small but non-vanishing width, and we will
model the situation with the following metric
\bea
ds^2=\frac1{z^2}\left(-f dv^2-2 R~ dv dz + d{\vec{x}}^2\right) \, ,
\label{g.global}
\eea
where
\bea
f&=&1-2m(v,z)~(R z)^d \, , \nonumber \\
R&=&R_0- (R_0-R_f)\frac {m(v,z)}M \, ,
\\m&=&\frac{M}{2}\left[1+\tanh \frac{w(v,z)}{w_0}\right] \, ,
\nonumber
\eea
being $w_0$ the parameter representing the thickness of the shell,
while $w(v,z)=0$ is the equation defining the position of the shell
in the $(v,z)$-plane. It is useful to define the quantities
\bea
\bar f&=& f(z,w=0) ~~~=~ 1-  M (\bar R z)^d,\cr\cr \bar R&=&
R(z,w=0) ~~~=~ \frac{R_0+R_f}2 \, .
\eea

By comparing the induced metric (\ref{gshell}) with equation
(\ref{g.global}) one finds $d\tau^2=\bar f dv^2+2\bar R~ dv~dz$.
Then, the position $\{v(\tau), z(\tau)\}$ of the shell satisfies the
equation
\bea
\bar f dv= \left( \frac{\sqrt{\bar f + \bar R^2\dot z ^2}}{\dot
z}-\bar R \right) dz \, ,
\eea
from which we find the equation describing the dynamics of the shell
to be $w=0$, with
\bea
w=v-\bar R \int_{z_0}^{z} \, dz \, \bar f^{-1} \, \left(
\sqrt{\frac{\bar f~ \bar R^{-2}+\dot z^2}{\dot{z}^2}}-1\right) \, ,
\label{wv}
\eea
where we set the following initial conditions $z(\tau_0)=z_0$ and
$v(\tau_0)=0$.

\subsection{Dynamical evolution of shells of matter}

Once the shell is at rest at $z=z_0$, in principle it is not
guaranteed that it will always collapse. For instance, we may think
of the shell as composed not by ordinary matter like baryons or
photons, but instead by the energy of a domain wall of a bubble
which encloses an AdS space in the interior with a given
cosmological constant, and an another AdS space in the outer region,
with a different one. Thus, it may occur that depending on the
values of the cosmological constants of the inner and outer regions
the bubble may collapse or expand.

Therefore, the dynamics of the shell depends of the sign of
$\frac{d\dot z^2}{dz}$ at $z_0$. It can be computed from equation
(\ref{zdot2}) and the result is
\bea
\left.\frac{d\,\dot z^2}{dz}\right|_{z_0}=\lambda(z_0) \, \xi(z_0)
\, ,
\eea
where
\bea
\lambda(z_0)=\frac{h(z_0)+R_0^{-2}-f_{out}(z_0) R_f^{-2}}{4h(z_0)}>0
\, .
\eea
The positivity follows from the positive energy condition. On the
other hand, $\xi(z_0)$ is defined as
\bea
\xi(z_0)=-(R_0^{-2}-R_f^{-2}) \frac{A}{z_0} +
A\left[R_0^{-1}-\sqrt{f_{out}(z_0)} R_f^{-1}\right]^2 z_0^{A-1} +
(d-A) R_f z_0^{d-1} \, .
\eea
Notice that in the cases of interest $0\leq A\leq d$. Hence, the
first term in the equation above is negative while the second and
third ones are positive. The conclusion is that in the case of equal
radii $\xi(z_0)>0$ and, therefore, the shell always collapses.

On the other hand, when $A=0$ (corresponding to $a=-1$) $\xi(z_0)$
is again positive and so the shell collapses independently of the
values $R_0$ and $R_f$.

The situation changes dramatically when we consider $A\neq0$ because
in the $z_0\rightarrow0$ limit the leading term is the first one
which is negative, implying that the shell generically expands.

For a given cut-off $z_0$ it can always be possible to find certain
radii $R_f>R_0$ such that $\xi(z_0)>0$. In order to observe collapse
the first term must be smaller than the others, therefore $R_0$ and
$R_f$ must be as close as possible. So, let us define $r=R_f-R_0<<1$
and for simplicity consider $R_0=1$. Therefore we find $r<<
z_0~10^{-2(2d-1)}$ for conformal matter, $r<< z_0~10^{-(2d-1)}$ for
massive dust, and $r<< z_0~10^{-2d}$ for relativistic matter. In
order to give an idea of the orders of magnitude involved, for
instance in the case of conformal matter in $AdS_5$, by setting the
initial position of the shell at $z_0=10^{-2}$, we need that
$R_f-R_0<<10^{-16}$ for the shell to collapse and, the limiting case
$z_0\rightarrow0$ only allows the fluid with $a=-1$, {\it i.e.} a
scalar field case, to collapse.

The same conclusion holds even if we relax the initial condition
$\displaystyle\left.\dot{z} \right|_{z_0}=0$. Indeed let us suppose
that $\displaystyle\left.\dot{z} \right|_{z_0}>0$, and then
extrapolate the shell position backward in time. If we assume
$\dot{z}\neq0$ for all $z<z_0$, then as $z\rightarrow0$ the $l.h.s.$
of (\ref{diff.eq}) vanishes if and only if $R_0=R_f$, but the
$r.h.s.$ vanishes for $A>0$ ($a>-1$).

Then for a collapsing fluid with $a>-1$ and $R_0\neq R_f$ the shell
can not be extrapolated to $z\rightarrow0$. Its velocity must vanish
at a certain position $\tilde z_0<z_0$ and so, in order for the
shell to collapse, $r$ must be smaller than in the case with the
shell at rest at $z_0$.

\section{Holographic thermalization}

In this section we describe the idea of holographic thermalization.
We will follow references \cite{Balasubramanian:2011ur} and
\cite{Galante:2012pv} and first consider two-point functions of
local gauge invariant QFT operators. For this purpose we look at
Wightman functions
\cite{Balasubramanian:2011ur,Balasubramanian:2010ce} of local gauge
invariant QFT operators ${\cal{O}}$ of conformal dimension $\Delta$.
We are interested in the equal time correlation functions. The point
is to study how these correlators change at different times.

On the other hand, using the gauge/string duality it is possible to
compute these correlators when the operators are heavy by using
geodesics in AdS spaces. We will compute the two-point functions
from a path integral as in references
\cite{Balasubramanian:1999zv,Balasubramanian:2011ur}
\begin{eqnarray}
< {\cal{O}} (t,\textbf{x}) {\cal{O}}(t, \textbf{x}')> = \int
{\cal{D}}{\cal{P}} e^{i \Delta L({\cal{P}})} \approx \sum_{
\textrm{geodesics}} e^{-\Delta {\cal{L}}} \, ,
\end{eqnarray}
where the path integral includes all possible paths connecting the
points at the AdS boundary, {\it i.e.} $(t, \textbf{x})$ and $(t,
\textbf{x}')$. In the above expression $L({\cal{P}})$ is the proper
length corresponding to this path. For space-like trajectories
$L({\cal{P}})$ is imaginary. The idea is to make a saddle-point
approximation for $\Delta \gg 1$. Therefore, only geodesics, {\it
i.e.} trajectories with extreme lengths will contribute. Notice that
in the last term ${\cal{L}}$ indicates actual length of the geodesic
between the points at the AdS boundary. In this way, there is a
direct relation between the logarithm of the equal-time two-point
function and the geodesic length between these two points. It is
important to be careful while considering these approximations
because the geodesic length diverges due to the AdS boundary
contributions. Then, one can define a renormalized distance $\delta
{\cal{L}} \equiv {\cal{L}} - 2 \ln (2/z_0)$, in terms of the cut-off
$z_0$ near the boundary, that suppresses the divergent part coming
from pure AdS.

The other type of non-local operators that we will be using are
spatial Wilson loops, which are non-local gauge invariant operators
in the field theory defined as the integral in a closed path
\textit{C} of the gauge field $A$. Wilson loops provide information
about the non-perturbative behavior of gauge theories, however, in
general it is difficult to compute them. Using the AdS/CFT
correspondence its computation can be done straightforwardly. The
expectation value of a Wilson loop is related to the string theory
partition function with a world-sheet $\Sigma$ extended on the bulk
interior, and ending on the closed contour \textit{C} on the
boundary,
\begin{eqnarray}
<W(C)> = < \frac{1}{N} Tr \left( {\cal{P}} e^{\oint_C A} \right)> =
\int {\cal{D}}\Sigma e^{-\Lambda(\Sigma)}  \simeq
e^{-\frac{1}{\alpha'}{\cal{A}}(\Sigma_0)} \, ,
\label{eqn_wilson_loops}
\end{eqnarray}
where, in the path integral one has to integrate over all the
non-equivalent surfaces whose boundary is  $\partial\Sigma=C$, at
the AdS boundary. $\Lambda(\Sigma)$ is the string action. The last
approximation in equation (\ref{eqn_wilson_loops}) is obtained in
the strong coupling regime by carrying out a saddle-point
approximation of the string theory partition function. In this way
we can reduce the computation of the expectation value of a Wilson
loop to determine the surface of minimal area of the classical
world-sheet whose boundary is \textit{C}. This will be a solution to
the equations of motion of the bosonic part of the string action
\cite{Maldacena:1998im,Rey:1998ik}.

These shell-collapsing models based on the AdS/CFT correspondence
allow to understand intuitively how the thermalization process takes
place. The outer region is described by a AdS-Schwarzschild black
hole, while the inner region is still an AdS space. Now, let us use
the geodesic approximation to compute the equal-time two-point
functions. If the separation of the boundary points is small enough,
then the geodesic cannot reach the shell at $w=0$ and, therefore,
the geodesic is seen as a purely AdS-Schwarzschild black hole
geodesic, {\it i.e.} for short distances in the field theory the
system seems to be in thermal equilibrium. If we increase the
separation between the insertion of the boundary operators, at some
point, the geodesic will cross the shell, and there will be a
geodesic refraction which will deviate it in comparison with the
thermal one. Thus, we can understand why the thermalization proceeds
from short to long distances, {\it i.e.} QFT ultraviolet degrees of
freedom thermalize first \cite{Balasubramanian:2011ur}.

In the next two subsections we discuss in more detail the
construction of the renormalized geodesic lengths and the
renormalized rectangular minimal area surfaces, which we will used
to probe thermalization of strongly coupled systems. These two
subsections are a generalization of our previous paper
\cite{Galante:2012pv} from where we follow the notation.

\subsection{Renormalized geodesics lengths}

In this subsection we focus on the evaluation of space-like geodesic
lengths as function of both time and boundary separation length.
Thus, we will consider space-like geodesics between points
$(t,x_1)=(t_0, -\ell/2)$ and $(t',x_1')=(t_0, \ell/2)$, where $\ell$
is the separation of the AdS boundary points. The orthogonal
coordinates are fixed. For instance, for $d=4$ we have
$(x_2,x_3)=(x_2',x_3')$. Therefore, we use as the geodesic parameter
the first coordinate $x_1$, that we simply call $x$. The solutions
to the geodesic equations are given by the functions $v(x)$ and
$z(x)$. Inserting a cut-off $z_0$ close to the AdS boundary, the
boundary conditions become
\begin{equation}
z(-\ell/2)  =  z_0 \, , \,\,\,\,\,\,\,\, z(\ell/2)   =  z_0 \, ,
\,\,\,\,\,\,\,\, v(-\ell/2) = t_0 \, , \,\,\,\,\,\,\,\,  v(\ell/2) =
t_0 \, .
\end{equation}
Also, $v(x)$ and $z(x)$ are symmetric under reflection $x
\rightarrow -x$. The geodesic length is defined as
\begin{eqnarray}
{\cal{L}}  =  \int \sqrt{-ds^2} = \int_{-\ell/2}^{\ell/2} dx
\frac{\sqrt{1-2R(v,z) z'(x)v'(x) - f(v,z) v'(x)^2}}{z(x)} \, ,
\label{ltermal}
\end{eqnarray}
where the prime indicates derivative with respect to $x$. Functions
$v(x)$ and $z(x)$ minimize the geodesic length of equation
(\ref{ltermal}). Since there is an $x$-independent Lagrangian, it
implies the existence of one conserved quantity, which is equivalent
to the Hamiltonian of the system. In terms of $f(v,z)$, the
conservation equation becomes
\begin{eqnarray}
1- 2 R(v,z) z' v' - f(z,v) v'^2 = \left( \frac{z_*}{z} \right)^2 \,
, \label{conservacion_two}
\end{eqnarray}
where the following initial conditions at the tip of the geodesic
have been used
\begin{eqnarray}
z(0)=z_* \, , \hspace{1.5cm} v(0)=v_* \, , \hspace{1.5cm} v'(0) =
z'(0) = 0 \, .
\end{eqnarray}

We can then solve the EOM for $v(x)$ and $z(x)$, obtaining
\begin{eqnarray}
0 & = & 1 -  v'(x)^2 f(v,z) - 2 R(v,z) v'(x) z'(x) -  R(v,z) z(x) v''(x) \nonumber\\
& & +\frac12 z(x) v'(x)^2 \partial_z f(v,z) - \frac12\partial_v R(v,z) z(x) v'(x)^2\, , \label{t_eom1} \\
0 & = & v''(x) f(v,z) + R(v,z) z''(x) + z'(x) v'(x) \partial_z f(v,z) \nonumber\\
& & + \frac{1}{2} v'(x)^2 \partial_v f(v,z) + \partial_z R z'(x)^2\,
, \label{t_eom2}
\end{eqnarray}
so we can just use these equations (and the conservation relation)
and replace $f(v,z)$ by the ones of interest to this work. Note that
for the different radii case, not only mass derivatives will appear
but also radius derivatives.

In order to evaluate the geodesic length as a function of $t_0$ and
the boundary separation $\ell$ we use the boundary conditions
\begin{eqnarray}
z(\ell/2)=z_0 \, ,  \hspace{1.5cm} v(\ell/2)=t_0 \, .
\end{eqnarray}
Now, the conservation equation and reflection symmetry lead to the
on-shell geodesic length given by the following expression
\begin{eqnarray}
{\cal{L}} (\ell, t_0) = 2 \int_{0}^{\ell/2} dx \frac{z_*}{z(x)^2} \,
,
\end{eqnarray}
Then, we must cancel the divergent part: $\delta {\cal{L}} (\ell,
t_0) = {\cal{L}} (\ell, t_0) - 2 \ln (2/z_0)$.

Thus, we can calculate how the thermalization process occurs by
considering a collapsing thin shell composed by different kind of
degrees of freedom. At this point we can start studying numerically
the thermalization process, by solving the EOM for different
starting $(v_*,z_*)$ values. We set the event horizon of the
thermalized geometry to be located at a position such that we have
always the same temperature at the final state. The results are
discussed in the next section, but before we introduce the formulas
of rectangular Wilson loops.

\subsection{Renormalized rectangular minimal area surfaces}

Now we carry out the computation of the minimal area surfaces. Using
the AdS metric with a shell, the Nambu-Goto action becomes,
\begin{eqnarray}
{\cal{A}}_{NG}(t_0,\ell,R_{WL})= \frac{R_{WL}}{2\pi}
\int^{\ell/2}_{-\ell/2} dx \frac{\sqrt{1-f (v,z) v'^2-2 R(v,z)
z'v'}}{z^2} \, , \label{area_rec_wil_loop}
\end{eqnarray}
for boundary rectangles parametrized by the coordinates $(x_1,x_2)$.
The rest of the coordinates at the AdS boundary are kept fixed. One
assumes the translational invariance along $x_2$. Then, we will use
$x_1$ to parametrize the functions $v(x_1)$ and $z(x_1)$ in the
AdS$_{d+1}$, and we call it $x$. Along the $x_2$ direction the
rectangular path on the boundary has length $R_{WL}$.

As in the previous case, there is no explicit dependence on $x$ and
therefore, there is a conserved quantity corresponding to the
Hamiltonian. The tip of the surface is $z_*$, with $z'(0) = v'(0) =
0$. Then, the conservation equation becomes
\begin{eqnarray}
1 - 2 R(v,z) z' v' - f(v,z) v'^2 = \left( \frac{z_*}{z} \right)^4.
\label{ec_cons_rwl}
\end{eqnarray}

The boundary conditions continue to be the same as in the geodesics
case,
\begin{equation}
z(-\ell/2)  =  z_0 \, , \,\,\,\,\,\,\,\, z(\ell/2)   =  z_0 \, ,
\,\,\,\,\,\,\,\, v(-\ell/2)  =  t_0 \, , \,\,\,\,\,\,\,\, v(\ell/2)
=  t_0 \, .
\label{bdary_cond}
\end{equation}

Next, we have to minimize the Nambu-Goto action for this geometry.
For our set up, these equations become
\begin{eqnarray}
0 & = & 2 - 2 f v'^2 - 4 R z' v' - R z v'' +   \frac{1}{2} z v'^2
\partial_z f
- (z v'^2 \partial_v R + \frac{1}{2}  z z'v' \partial_z R) \\
0 & = & 2 f^2 v'^2 - f (2 - 4 R v' z' + \frac12 z v'^2 \partial_z f)
- z ( R^2 z'' +  R \partial_z f v' z' +  \nonumber \\
& & +\frac12 R v'^2 \partial_v f) +  z (f v'^2 \partial_v R - R z'^2
\partial_z R)
\end{eqnarray}

We can again extract the physical information of time and boundary
separation length from the boundary conditions (\ref{bdary_cond})
and rewrite the on-shell Nambu-Goto action by making use of the
conservation equation, obtaining
\begin{eqnarray}
{\cal{A}}(t_0,\ell,R_{WL})= \frac{R_{WL}}{\pi} \int_{0}^{\ell/2} dx
\, \frac{z_*^2}{z^4} \, .
\end{eqnarray}
Finally, we subtract the divergent part from pure AdS space by
defining
\begin{eqnarray}
\delta {\cal{A}}(t_0,\ell) = \frac{\pi}{R_{WL}} \left(
{\cal{A}}(t_0,\ell,R_{WL}) - \frac{1}{z_0} \frac{R_{WL}}{\pi}
\right) \, .
\end{eqnarray}

Now, we focus on the results obtained by solving the differential
equations for both renormalized space-like geodesic lengths and
rectangular minimal area surfaces.

\section{Results of dynamical holographic thermalization}

In this section we introduce our results obtained from numerical
calculations, by solving the system of differential equations
described in the previous section for the evolution of thin shells,
using renormalized geodesic lengths and rectangular minimal area
surfaces as extended probes of thermalization of QFT strongly
coupled systems.

\begin{figure}
\centering \subfigure[AdS$_5$]{
\includegraphics[scale=0.8]{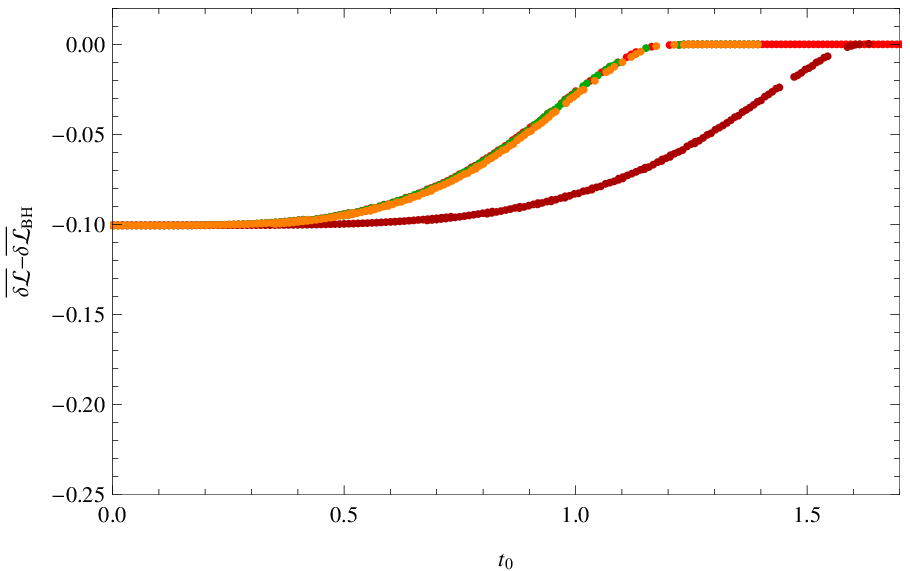}
\label{figure1a} } \subfigure[AdS$_5$]{
\includegraphics[scale=0.8]{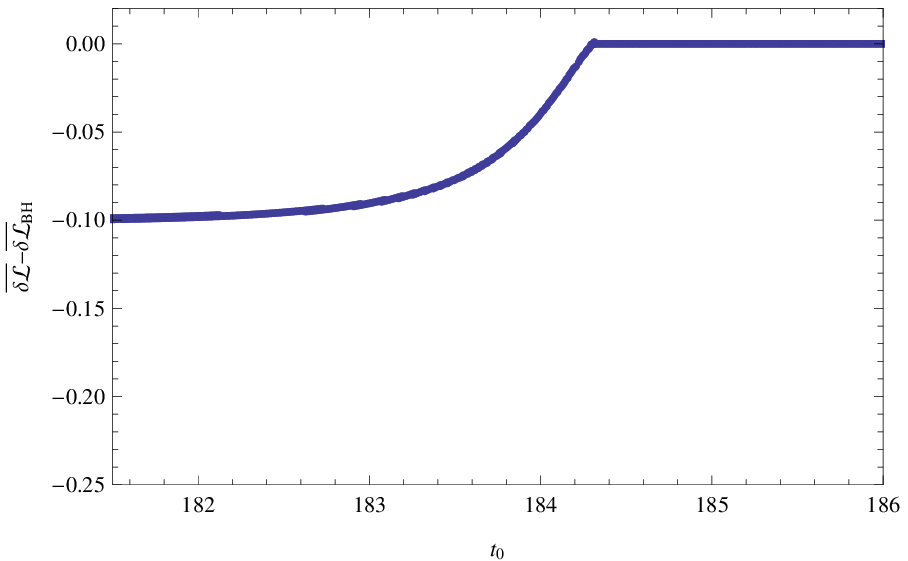}
\label{figure1b} }
\subfigure[AdS$_4$]{
\includegraphics[scale=0.8]{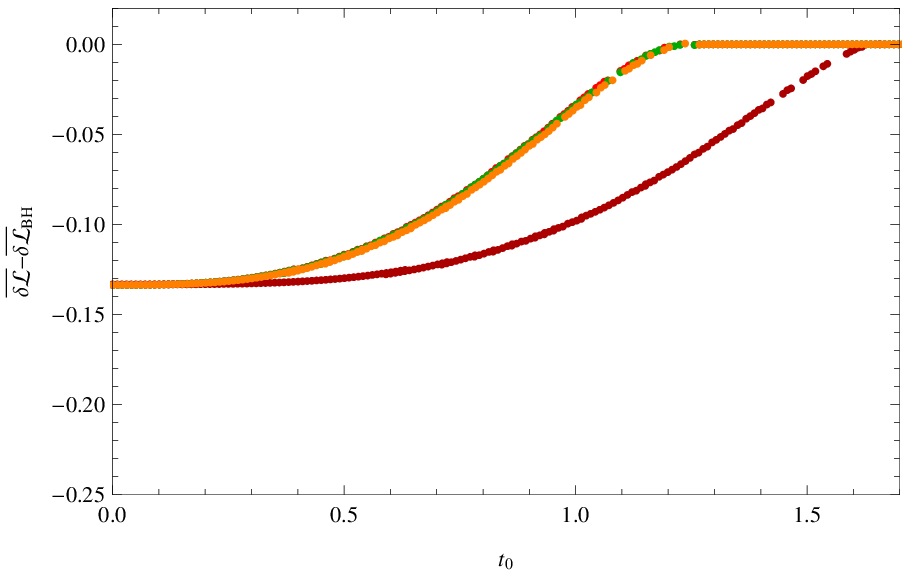}
\label{figure2a} } \subfigure[AdS$_4$]{
\includegraphics[scale=0.8]{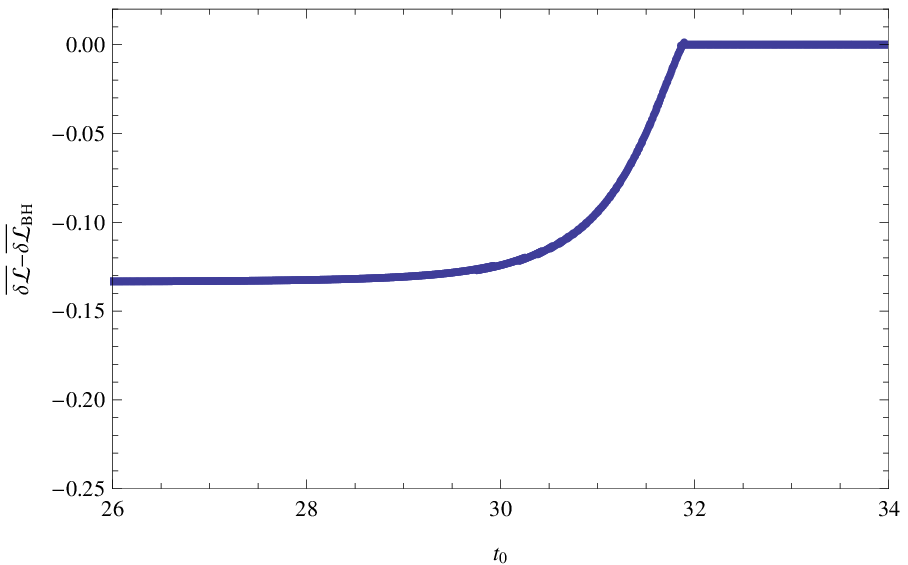}
\label{figure2b} }
\subfigure[AdS$_3$]{
\includegraphics[scale=0.8]{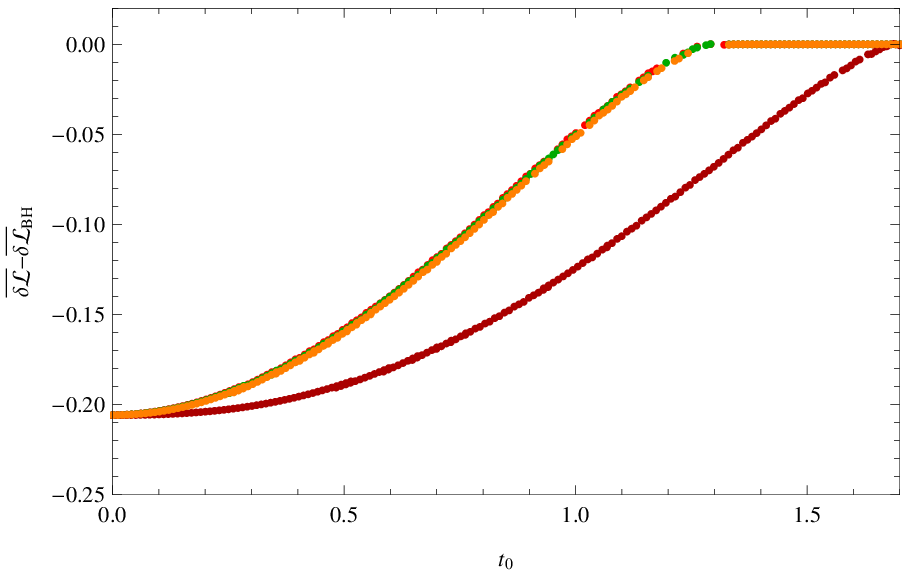}
\label{figure3a} } \subfigure[AdS$_3$]{
\includegraphics[scale=0.8]{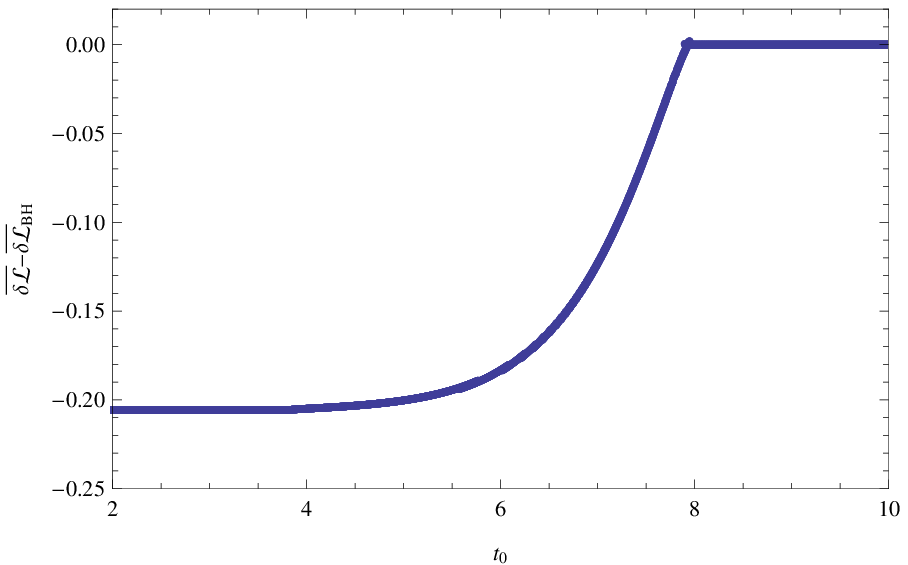}
\label{figure3b} }
\caption{\small Thermalization of the renormalized space-like
geodesic lengths for the boundary separation $\ell=2.6$, considering
$R_0=R_f=1$, for boundary theory dimensions $d=4$, 3 and 2,
indicated as AdS$_5$, AdS$_4$ and AdS$_3$, respectively. In each
figure curves for different matter are indicated with different
colors: for AdS-Vaidya (red curve), scalar field (green curve),
massive dust (orange curve), relativistic matter (dark red curve),
conformal matter (blue curve). Following the literature we plot the
difference between the geodesic length and the thermal geodesic
length divided by the boundary separation $\ell$. The same applies
for the rest of the figures.} \label{f1gl}
\end{figure}

First, in figure \ref{f1gl} we show the results for thermalization
of the renormalized space-like geodesic lengths for the boundary
separation $\ell=2.6$, by considering $R_0=R_f=1$ and $2 M=1$, for
the boundary QFT theory dimensions $d=4$, 3 and 2, indicated as
AdS$_5$, AdS$_4$ and AdS$_3$, respectively. The cases with a shell
composed of a scalar field (green curve) and dust (orange curve),
both almost coincide with the Vaidya shell (red curve) as it is
shown in figures 1.a, 1.c and 1.e. It turns out that the Vaidya
shell thermalizes first. Slightly later it does the shell composed
by a scalar field and then, almost at the same time the shell of
dust. In fact, these three cases depicted in figures 1.a, 1.c and
1.e, almost completely overlap. In the same figure the dark-red
curve indicates relativistic matter, which thermalizes later. Notice
that for relativistic matter, whose EOS has $a=c/(d-1)$, we have the
freedom to set $0<c<1$, being the pressureless and conformal matter
the limiting cases. As $c$ increases, so does the thermalization
time for relativistic matter, approaching the conformal matter time
scale. Figures 1.b, 1.d and 1.f show thermalization when considering
conformal matter (blue curve), which occurs at $t_0$ much larger
than the other cases. The fact that AdS-Vaidya, scalar field and
dust shells coincide is a general result which does not depend on
boundary separation. This is so because the integrand on the {\it
r.h.s.} of equation (\ref{wv}) is much smaller than one for any
value of $z$. Thus, the equation describing the position of the
shell is $v \simeq 0$, as in the AdS-Vaidya case. The larger
thermalization time found for the conformal matter case is closely
related to the fact that the {\it r.h.s.} of equation (\ref{wv})
takes a non-zero asymptotic value for large $z$. This value
increases with space-time dimension, making conformal matter in
higher dimensions to thermalize later.

Another interesting possibility is to consider $R_0=R_f$ with
different values. In fact, we have considered $R_0=R_f=0.5$ in
figures 2.a and 2.b and $R_0=R_f=2$ in figures 2.c and 2.d. These
cases are for systems going from AdS$_5$ to an AdS$_5$-Schwarzschild
black hole. In both cases we can see that AdS-Vaidya (red curve), a
massive dust (orange curve) and a scalar field (green curve)
thermalize almost simultaneously, relativistic matter does it a bit
later (dark-red curve), and much later conformal matter (blue
curve). This difference can be better appreciated from the insets of
both figures. In all these curves we keep the dimensionless product
of the boundary separation length by the plasma equilibrium
temperature $\ell \, T$ fixed, thus by changing $R_f$ the horizon
changes as $z_h=1/R_f$.
\begin{figure}
\centering \subfigure[AdS$_5$, $R_0=R_f=0.5$]{
\includegraphics[scale=0.5]{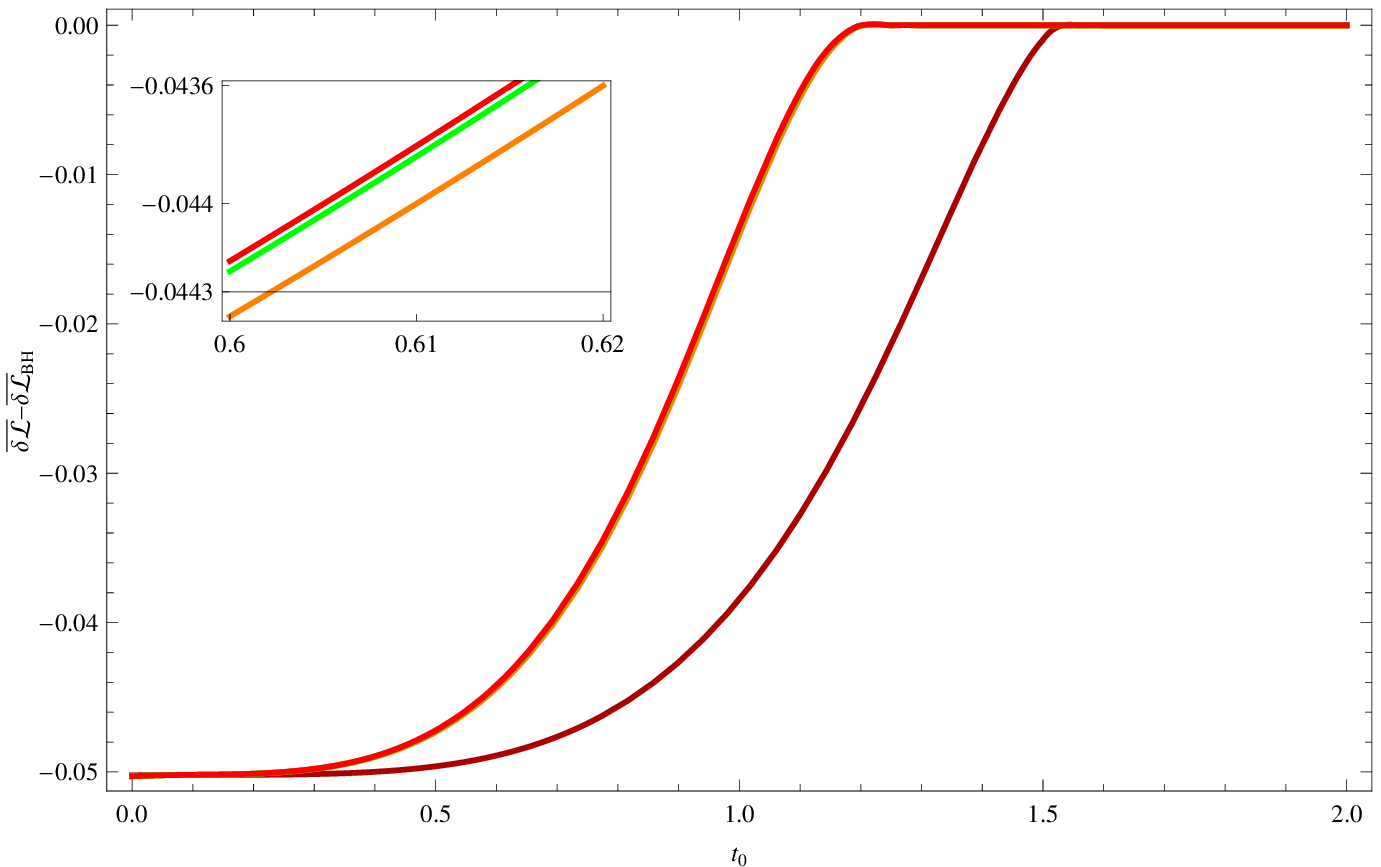}
\label{figure} } \subfigure[AdS$_5$, $R_0=R_f=0.5$]{
\includegraphics[scale=0.5]{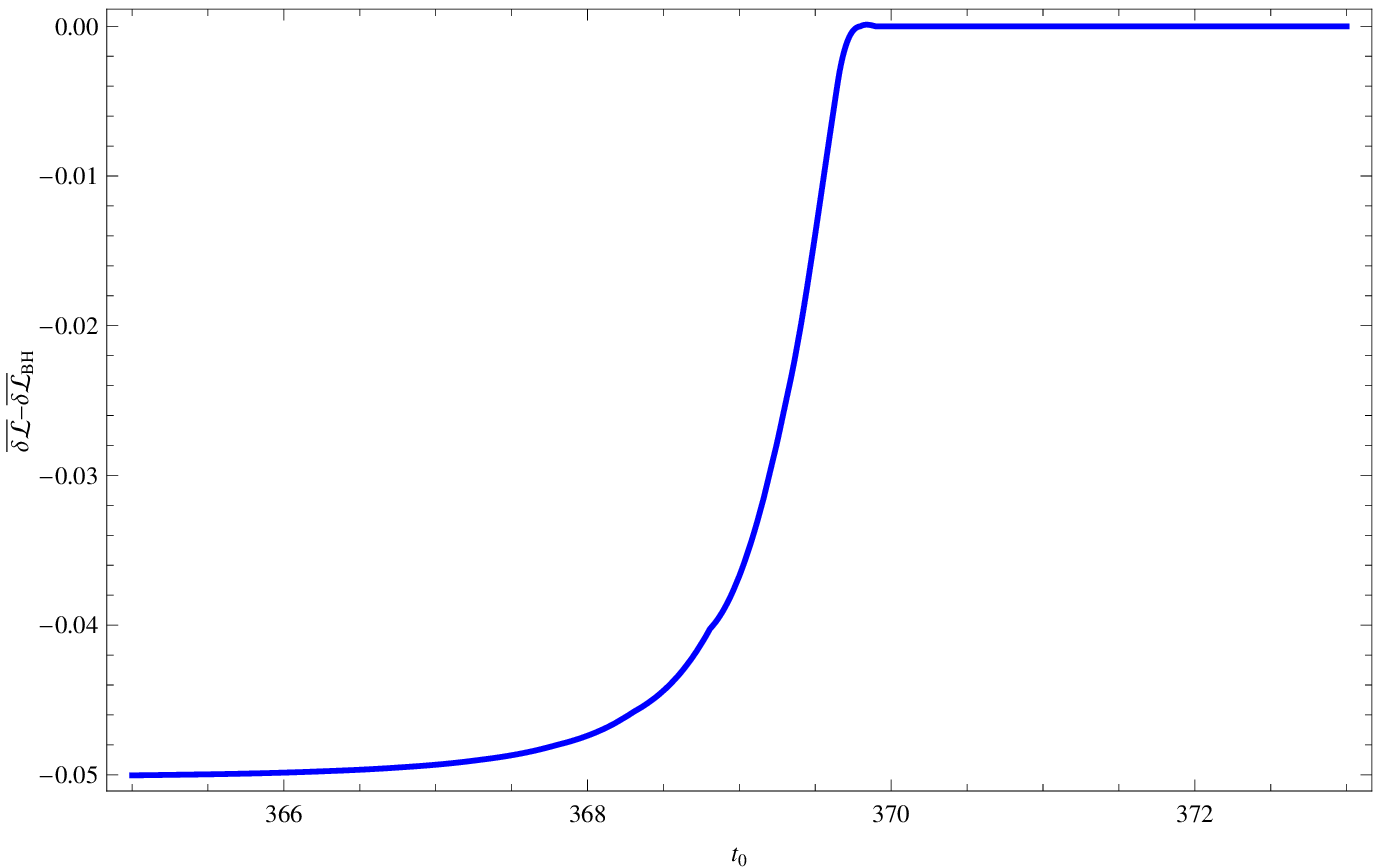}
\label{figure Rneq05} }
\centering \subfigure[AdS$_5$, $R_0=R_f=2$]{
\includegraphics[scale=0.5]{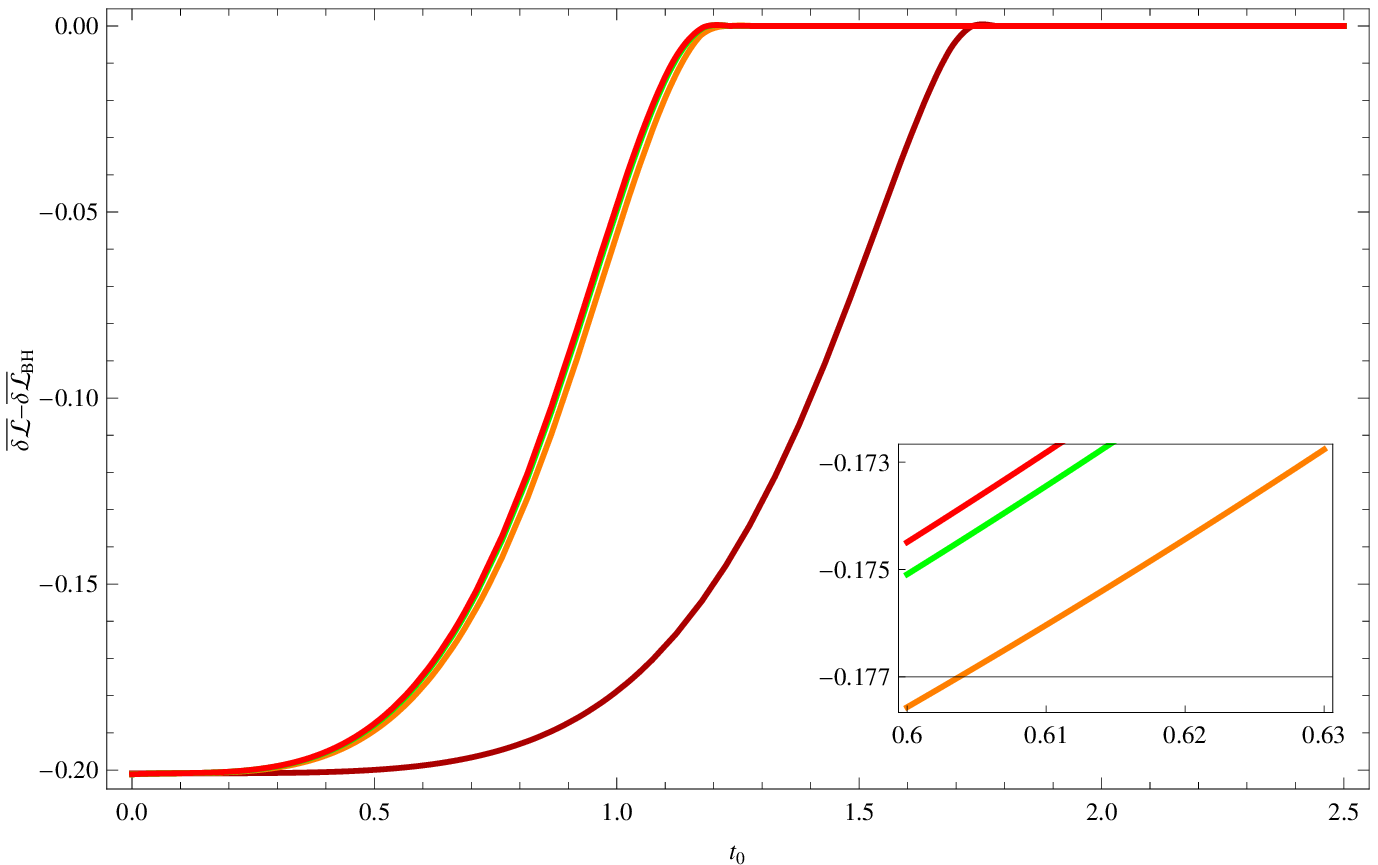}
\label{figure} } \subfigure[AdS$_5$, $R_0=R_f=2$]{
\includegraphics[scale=0.5]{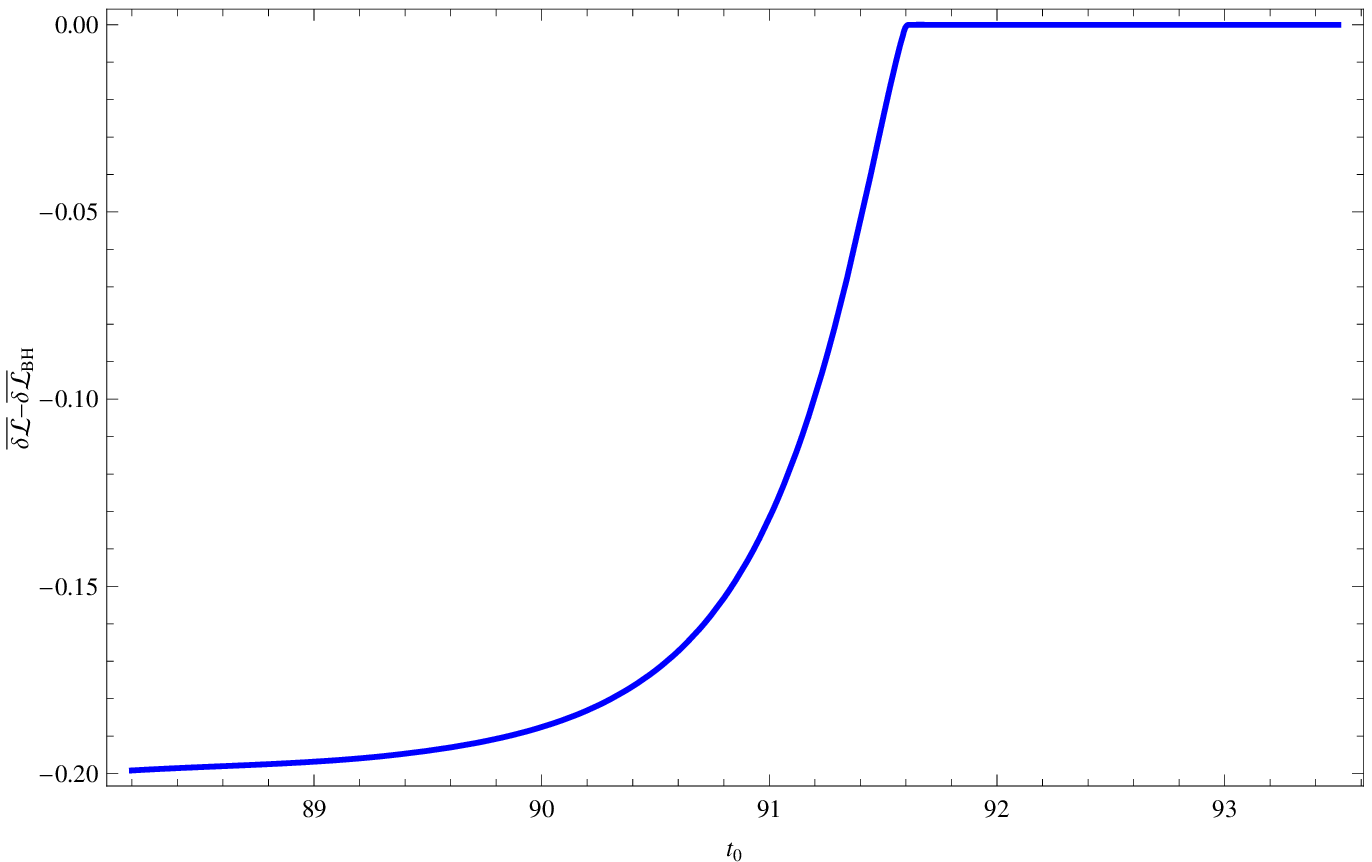}
\label{figure Rneq2} }
\caption{Renormalized space-like geodesic length as a function of
time for Vaidya-type (red curve), scalar field (green curve),
massive dust (orange curve), relativistic matter (dark red curve),
conformal matter (blue curve) shells, respectively. Insets zoom in
the first curve in both figures.}
\end{figure}

We can also make a comparison between figures 1 and 2.  For the
AdS-Vaidya, massive dust, and a scalar field cases the
thermalization time scale is not sensitive to the changes of the
radii, in the range considered, {\it i.e.} $R_0=R_f=0.5$, 1 and 2.
The more remarkable effect is that for conformal matter where for
$R_0=R_f=0.5$, 1 and 2, the thermalization time decreases
notoriously, as can be seen from the figures. In the particular case
of relativistic matter considered we observe a small enhancement of
the thermalization time as the radii increase, but of course the
thermalization of this kind of matter strongly depends on the value
$a$.

Another situation that we have investigated is the case when the
inner and outer radii are different. As it has been explained
before, only a shell composed by a scalar field can thermalize in
this case. In order to illustrate the behavior we have considered:
$R_0=0.5$, while $R_f=1$ (red curve), $R_f=2$ (blue curve). For
these cases the thermalization time is $t_0 \approx 90$ and 60,
respectively, while $\overline{\delta {\cal {L}}}-\overline{\delta
{\cal {L}}_{BH}}=-2$, $-5$. This is shown in figure 3. By
incrementing the difference between $R_0$ and $R_f$ it is possible
to recover short thermalization times, for instance the case with
$R_0=0.5,~R_f=10$ gives $t_0\approx 10$ (not displayed here). Notice
that the thermalization scales are not monotonous with respect to
the difference between both radii. This is so because by changing
the radii one varies the velocity of the shell as well as the
position $z_*$ of the thermalized geodesic tip.
\begin{figure}
\centering {
\includegraphics[scale=0.8]{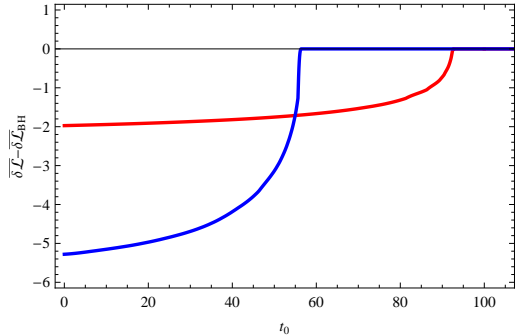}
\label{R0neqRf}} \caption{Renormalized geodesic length differences
when $R_0=0.5$; $R_f=1$ (red curve) and $R_f=2$ (blue curve).}
\end{figure}

\begin{figure}
\centering \subfigure[AdS$_5$]{
\includegraphics[scale=0.8]{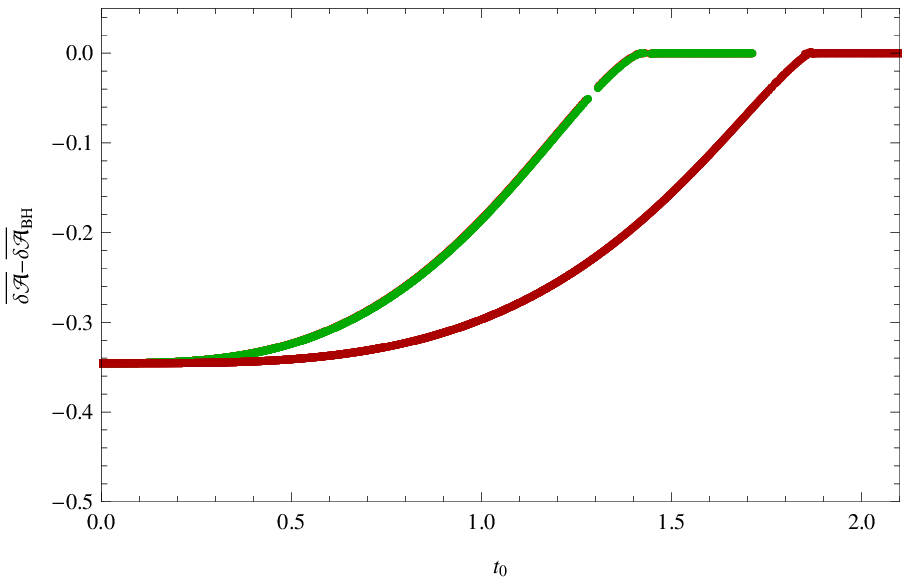}
\label{figure1a} } \subfigure[AdS$_5$]{
\includegraphics[scale=0.8]{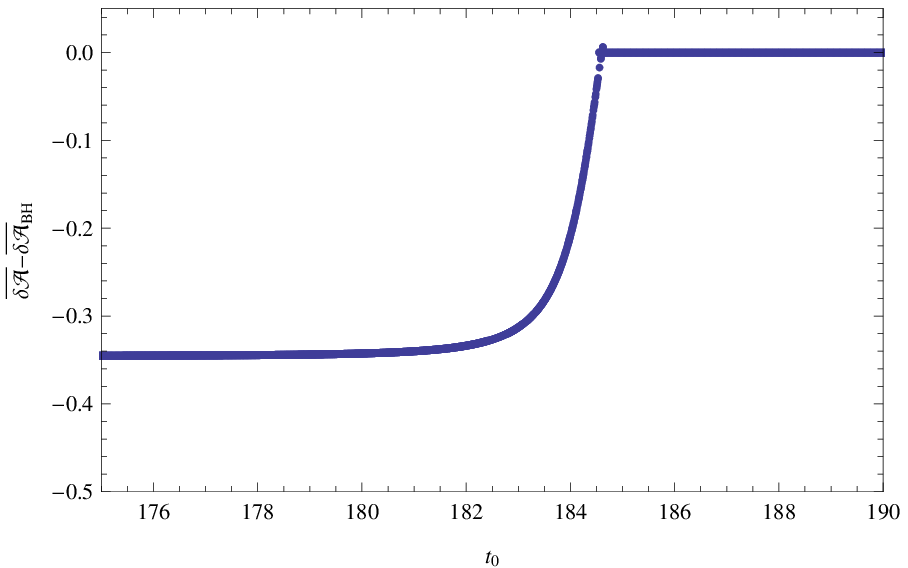}
\label{figure1b} }
\subfigure[AdS$_4$]{
\includegraphics[scale=0.8]{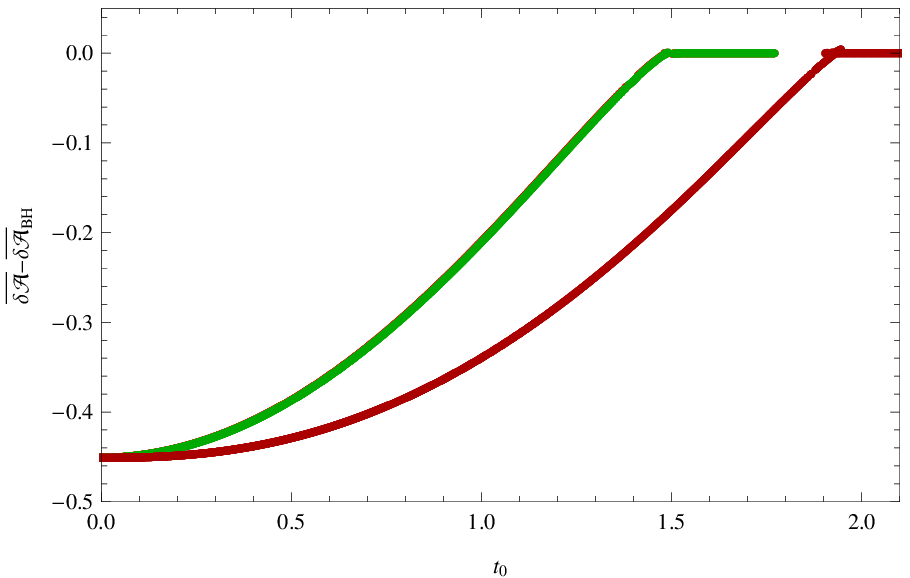}
\label{figure2a} } \subfigure[AdS$_4$]{
\includegraphics[scale=0.8]{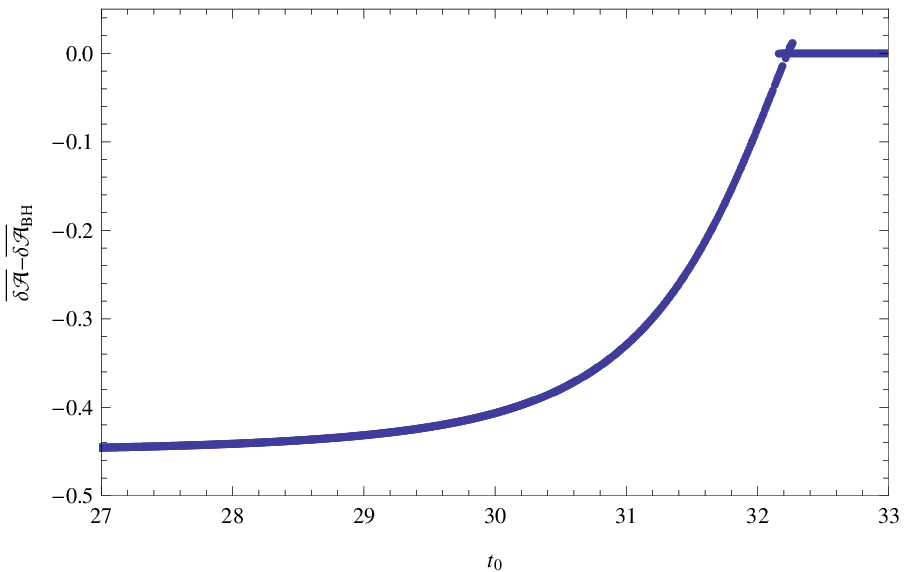}
\label{figure2b} }
\caption{\small Thermalization of the renormalized minimal area
surfaces, with $\ell=2$, considering $R_0=R_f=1$, for the boundary
theory dimensions $d=4$ and 3, labeled by AdS$_5$ and AdS$_4$,
respectively. Different kinds of matter in the shell are indicated
by colored curves as follows: Vaidya-type (red curve), scalar field
(green curve), massive dust (orange curve), relativistic matter
(dark red curve), conformal matter (blue curve) shells.} \label{fwl}
\end{figure}

Figure \ref{fwl}, on the other hand, shows a similar behavior as
figure 1 for renormalized rectangular minimal area surfaces for
$d=3$ and 4. In this case we also set $R_0=R_f=1$ and $2 M=1$. We
observe the same trend as in figure 1. The thermalization shown in
figure 4 corresponding to rectangular Wilson loops in the dual QFT
shows the appearance of swallow tails when thermal equilibrium is
reached. Something similar was observed before in the case of an
AdS-Vaidya shell \cite{Balasubramanian:2011ur}, and even in the
cases with an AdS-Vaidya shell composed by charged dust
\cite{Galante:2012pv}.
%

\section{Discussion and conclusions}

In this paper we have studied dynamical evolution of thin shells
composed by different degrees of freedom in AdS spaces, obtaining
different thermalization time scales. We have used the thin-shell
formalism, applying the Israel junction conditions, and also imposed
the positive energy conditions. Thus, we obtain a general framework
where the distinction in the composition of the shells is made
explicit through the equation of state in each case. We have also
explored different space-time dimensions.

We have considered an AdS-Vaidya shell, which can be understood as
composed by massless dust, moving at the speed of light, and then we
also investigated shells made of a scalar field, a pressureless
massive fluid, the so-called relativistic matter, and matter whose
energy-momentum tensor is traceless. The parameters to play with are
the space-time dimension $d$, and the radii of the inner and outer
regions, $R_0$ and $R_f$.

The first observation is that when the $R_0=R_f$, the thermalization
time scales of the AdS-Vaidya, the scalar field and a pressureless
massive fluid shells, are the same. The conformal case thermalizes
much later, strongly depending on space-time dimensions and radii.
Relativistic matter case continuously interpolates between both
cases, depending on its EOS.

In addition, we have studied the effect on the thermalization curves
when the inner and outer radii are different. Also, we have found
that the positive energy condition implies that the inner radius
must be equal or smaller than the outer one, which means that the
absolute value of the vacuum energy density of the inner region must
be equal or larger than the one of the outer region. Finally, we
have found that only in the case of a shell composed by a scalar
field the positive energy condition allows for the shell to
collapse.

When the energy densities of the inner and outer spaces differ, the
thermalization time scales considerably increases. For instance, for
the scalar field case, which for equal radii coincides with the
AdS-Vaidya shell, for different radii the thermalization time can be
set arbitrarily large. Some particular examples where displayed in
figure 3.

The main conclusion from this work is that holographic models do not
necessarily yield a rapid thermalization. Moreover, the
thermalization time scale strongly depends on the equation of state
governing the shell. This will determine the shell velocity and
consequently, thermalization times. We show that it is possible to
have EOS that lead to delayed thermalization times (such as the case
of conformal matter).

There are other possible directions where the ideas and formalism
presented here can be extended. For instance, while changing the
composition of the shell we will be imposing different shell
velocities. This allows one to model different possible scenarios
for the evolution of thermalization processes in strongly coupled
systems. One is to consider lower-dimensional systems in the context
of AdS/CMT. Another aspect concerns the study of a quantum quench
across critical points \cite{Calabrese:2006rx,Calabrese:2007rg}. For
example a quantum quench across a zero temperature holographic
superfluid transition has recently been reported in
\cite{Basu:2012gg}. Another very interesting extension could be
along the lines of the recent work by Buchel, Lehner and Myers,
where it has been studied thermal quenches in a particular mass
deformation of the ${\cal {N}} = 4$ SYM theory. There is a
transition between an initial thermal state of ${\cal {N}} = 4$ SYM,
to a final state with the mentioned mass deformation which yields
the so-called ${\cal {N}} = 2^*$ SYM theory. This transition has
been described in terms of a thermal quench \cite{Buchel:2012gw}.

~

\centerline{\large{\bf Acknowledgments}}

~

We thank Alex Buchel, Johanna Erdmenger, Nicol\'as Grandi, Luis
Lehner, Shu Lin, Juan Maldacena, Carlos N\'u\~nez, Guillermo Silva
and Gianmassimo Tasinato for useful discussions and comments. We
specially thank Alex Buchel for a critical reading of the
manuscript. The work of W.B. and M.S. has been supported by CONICET,
the Consejo Nacional de Investigaciones Cient\'{\i}ficas y
T\'ecnicas of Argentina, and the ANPCyT-FONCyT Grant
PICT-2007-00849. The work of M.S. has also been supported by the
CONICET Grant PIP-2010-0396. Research at Perimeter Institute is
supported by the Government of Canada through Industry Canada and by
the Province of Ontario through the Ministry of Research $\&$
Innovation. D.G. also acknowledges support from an NSERC Discovery
grant.

\end{document}